\documentclass[a4paper,twocolumn,unpublished]{quantumarticle}
\pdfoutput=1
\usepackage[utf8]{inputenc}
\usepackage[T1]{fontenc}

\usepackage{amsthm}
\usepackage{amsmath}
\usepackage{amssymb}
\usepackage{mathtools}
\usepackage{graphicx}
\usepackage{hyperref}
\usepackage[numbers,sort&compress]{natbib}

\theoremstyle{definition}
\newtheorem{definition}{Definition}
\newtheorem{remark}{Remark}
\newtheorem{result}{Result}
\newtheorem{problem}{Problem}
\newtheorem{possibility}{Possibility}

\DeclareMathOperator\tr{tr}
\DeclarePairedDelimiter\abs\lvert\rvert

\usepackage{xcolor}
\usepackage{comment}

\begin{document}


\title{Possible consequences for physics of the negative resolution of Tsirelson's problem}


\author{Ad\'an Cabello}
\email{adan@us.es}
\affiliation{
Departamento de F\'{\i}sica Aplicada II,
Universidad de Sevilla,
41012 Sevilla,
Spain
}
\affiliation{
Instituto Carlos~I de F\'{\i}sica Te\'orica y Computacional,
Universidad de Sevilla,
41012 Sevilla,
Spain
}
\orcid{0000-0002-4631-457X}

\author{Marco Túlio Quintino}
\email{Marco.Quintino@lip6.fr}
\affiliation{Sorbonne Universit{\' e}, CNRS, LIP6, F-75005 Paris, France}
\orcid{0000-0003-1332-3477}

\author{Matthias Kleinmann}
\email{matthias.kleinmann@uni-siegen.de}
\affiliation{Naturwissenschaftlich-Technische Fakultät, Universität Siegen, Walter-Flex-Straße 3, 57068 Siegen, Germany}
\orcid{0000-0002-5782-804X}


\begin{abstract}
In 2020, Ji \emph{et al.} [arXiv:2001.04383 and Comm.~ACM \textbf{64}, 131 (2021)] provided a proof that the complexity classes $\text{MIP}^\ast$ and $\text{RE}$ are equivalent. This result implies a negative resolution of Tsirelson's problem, that is, $C_{qa}$ (the closure of the set of tensor product correlations) and $C_{qc}$ (the set of commuting correlations) can be separated by a hyperplane (that is, a Bell-like inequality). In particular, there are correlations produced by commuting measurements (a finite number of them and with a finite number of outcomes) on an infinite-dimensional quantum system which cannot be approximated by sequences of finite-dimensional tensor product correlations.
Here, we point out that there are four logical possibilities of this result. Each possibility is interesting because it fundamentally challenges the nature of spacially separated systems in different ways.
We list open problems for making progress for deciding which of the possibilities is correct.
\end{abstract}


\maketitle


\section{Definitions}


The result of Ji \emph{et al.} \cite{ji2022mipre,Ji:2021CACM} on the complexity classes $\text{MIP}^\ast$ and $\text{RE}$ has implications regarding the relation between two sets of quantum correlations. It is therefore appropriate to start by providing definitions of the relevant sets of correlations.

\begin{definition}[Correlation]
For a bipartite Bell scenario $(\abs X,\abs Y,\abs A,\abs B)$ \cite{Bell:1964PHY,Clauser:1969PRL}, in which all of Alice's measurements $x \in X$ have the same outcome set $A$, and all of Bob's measurements $y \in Y$ have the same outcome set $B$, and $A,B,X,Y$ are finite sets, a \emph{correlation} (this is the term used in, for example, \cite{Wright:2022ARX}), or \emph{correlation matrix} (term used in, for example, \cite{Slofstra:2019FMP}), or \emph{behavior} \cite{Tsirelson:1993HJS}, or \emph{empirical model} \cite{Abramsky:2011NJP}, or \emph{probability model} \cite{Abramsky:2012PRA}, or \emph{box} (term used in, for example, \cite{Ramanathan:2016PRL}), is the collection $p(a,b|x,y)$ for all $x \in X$ and $y \in Y$, and all $a \in A$ and $b \in B$. That is, it is a list of probability distributions, one for each pair $(x,y)$.
\end{definition}

Note that here and in the following the sets $A$, $B$, $X$, $Y$ are arbitrary, but fixed.

\begin{definition}[The set of quantum correlations $C_{q}$ \cite{paulsen2013quantumchromaticnumbersoperator,Paulsen2014Estimating,Slofstra2016Tsirelson}] \label{Cq}
A correlation $p$ is in the set $C_{q}$ if there are separable Hilbert spaces ${\cal H}_A$ and ${\cal H}_B$ of finite dimension, positive operator-valued measures (POVMs) (which, in the finite-outcome case, is a collection of positive semidefinite operators summing to identity) $A^x = \{A^x_a\}_{a \in A}$ for all $x \in X$ on ${\cal H}_A$ and $B^y = \{B^y_b\}_{b \in B}$ for all $y \in Y$ on ${\cal H}_B$, and a density operator (that is, a positive semidefinite operator with unit trace) $\rho$ on ${\cal H}_A \otimes {\cal H}_B$ such that
\begin{equation}
\label{equ}
p(a, b|x, y) = \tr (A^x_a \otimes B^y_b \rho).
\end{equation}
\end{definition}

\begin{definition}[The set of quantum-spatial correlations $C_{qs}$ \cite{paulsen2013quantumchromaticnumbersoperator,Paulsen2014Estimating,Slofstra2016Tsirelson}] \label{Cqs}
The same as Definition~\ref{Cq}, but replacing $C_q$ by $C_{qs}$ and ``of finite dimension'' by ``(possibly infinite dimensional)''.
\end{definition}

\begin{definition}[The set of quantum-approximable correlations $C_{qa}$ \cite{paulsen2013quantumchromaticnumbersoperator,Paulsen2014Estimating,Slofstra2016Tsirelson}] \label{Cqa}
$C_{qa}$ is the closure of $C_{q}$. That is, all tensor product correlations that can be approximated arbitrarily well by finite dimensions.
\end{definition}

\begin{definition}[The set of quantum-commuting correlations $C_{qc}$ \cite{paulsen2013quantumchromaticnumbersoperator,Paulsen2014Estimating,Slofstra2016Tsirelson}] \label{Cqc}
A correlation $p$ is in the set $C_{qc}$ if there is a separable Hilbert space ${\cal H}$ (possibly infinite dimensional), POVMs $A^x = \{A^x_a\}_{a \in A}$ for all $x \in X$ on ${\cal H}$ and $B^y = \{B^y_b\}_{b \in B}$ for all $y \in Y$ on ${\cal H}$ such that $[A^x_a,B^y_b]=0$ for all $x,y,a,b$, and a density operator $\rho$ on ${\cal H}$ such that
\begin{equation}
\label{equ3}
p(a, b|x, y) = \tr (A^x_a B^y_b \rho),
\end{equation}
where $A^x_a B^y_b$ is the product of the operators $A^x_a$ and $B^y_b$.
\end{definition}

\begin{remark}
$C_{qs}$ is commonly used to model correlations between measurements on two spatially separated systems in non-relativistic quantum mechanics (QM) and non-relativistic quantum information \cite{CohenTannoudji1973, nielsen_chuang_2010}. 
\end{remark}

\begin{remark}
In algebraic quantum field theory (AQFT), local observables are represented by operators acting on a joint Hilbert space and $C_{qc}$ is, in principle, the set of correlations between experiments in spacelike separated regions \cite{Haag1992,Halvorson2007hand}.
However, extra constraints are sometimes added to handle the infinitely many degrees of freedom. These constraints imply that the local algebras of strictly separated space-time regions are contained in Hilbert space tensor factors. This is called the ``split property'' \cite{Doplicher1984IM}. Therefore, $C_{qs}$ would be the set of correlations between spacelike separated regions satisfying this split property. However, ``if an appropriate correlation expression could be constructed and implemented in the laboratory, the split property could be refuted experimentally'' \cite{Junge:2011JMP}.
\end{remark}

\begin{remark}
$C_{qc}$ is also the way QM models correlations between mutually non-disturbing measurements of jointly measurable sharp observables. For example, those produced by sequential ideal measurements of jointly measurable sharp observables for Kochen-Specker contextuality \cite{Budroni:2022RMP}.
\end{remark}

\begin{remark}
The notation $C_q, C_{qs}, C_{qa}, C_{qc}$ dates back to Refs.~\cite{paulsen2013quantumchromaticnumbersoperator, Paulsen2014Estimating, Dykema2015Synchronous, Slofstra2016Tsirelson}. The definitions readily imply $C_q \subseteq C_{qs} \subseteq C_{qa} \subseteq C_{qc}$ and one finds that $C_{qc}$ is already closed~\cite{Paulsen2014Estimating, Slofstra2016Tsirelson}.
\end{remark}

\begin{definition}[Tsirelson's problem \cite{Scholz2008ARX}]
Is $C_{qa}$ equal to $C_{qc}$? Or, equivalently, can all infinite-dimensional commuting correlations be approximated by sequences of finite-dimensional tensor product correlations?
\end{definition}


\section{Summary of previous results}


\begin{result}
Scholz and Werner \cite{Scholz2008ARX} showed that $C_{qa} = C_{qc}$ holds under the assumption that Alice's or Bob's operator algebra is nuclear.
This is the case if, for instance, Alice is restricted to perform two different dichotomic measurements or if the underlying Hilbert space is finite dimensional \cite{TsirelsonWeb}.
\end{result}

\begin{result}
Junge \emph{et al.}~\cite{Junge:2011JMP} and Fritz~\cite{Fritz:2012RMP} showed that Tsirelson’s problem and Connes' embedding problem on finite approximations in von Neumann algebras (known to be equivalent to Kirchberg's QWEP conjecture) are equivalent.
\end{result}

\begin{result}
Coladangelo and Stark \cite{Coladangelo:2020NC} showed that, in the bipartite Bell scenario $(4,5,3,3)$ in which Alice (Bob) has $4$ ($5$) settings with $3$ outcomes, there exists a correlation which is not attainable in $C_q$ but it is attained in $C_{qs}$ by infinite-dimensional quantum systems, hence $C_{q} \neq C_{qs}$.
\end{result}

\begin{result}
Slofstra \cite{Slofstra:2019FMP} showed that $C_{qs} \neq C_{qa}$, that is, the set $C_{qs}$ is not closed. The proof is constructive, with the Bell scenario in question having input sets of size 184 and 235, and output sets of size 8 and 2. Hence the Bell scenario is $(184,235,8,2)$. Later, Dykema \emph{et al.} \cite{Dykema:2019CMP} showed that $C_{qs} \neq C_{qa}$ already in the $(5,5,2,2)$ scenario. This proof was then simplified in Ref.~\cite{Musat:2020CMP}.
\end{result}

\begin{result}
Ji \emph{et al.} \cite{ji2022mipre, Ji:2021CACM} showed the equivalence of the complexity classes $\text{MIP}^\ast$ and $\text{RE}$ and that this then implies $C_{qa} \neq C_{qc}$ for some Bell scenario with finite $n=\abs A=\abs B$ and $k=\abs X=\abs Y$. Hence, not all infinite-dimensional commuting correlations can be approximated by sequences of finite-dimensional tensor product correlations.

Another consequence of $\text{MIP}^\ast=\text{RE}$ \cite{ji2022mipre, Ji:2021CACM} is that the $\varepsilon$-weak membership problem for $C_{qa}$ is undecidable for some $\varepsilon>0$. That is, one cannot design a universal algorithm that decides if a correlation $p$ from an arbitrary Bell scenario is $\varepsilon$-close (in the $l_1$ distance) to the set $C_{qa}$.
\end{result}

\begin{remark}
The proof in Ref.~\cite{ji2022mipre} ``yields an explicit correlation that is in the set $C_{qc}$ but not in $C_{qa}$. [\ldots]
It is in principle possible to determine an upper bound on the parameters $n$ [number of measurement settings of both Alice and Bob, that is, $n=\abs X=\abs Y\,$], $k$ [number of outcomes of all measurements, that is, $k=\abs A=\abs B\,$] for our separating correlation from the proof. While we do not provide such a bound, there is no step in the proof that requires it to be astronomical; e.g.\ we believe (without proof) that $10^{20}$ is a clear upper bound'' \cite{ji2022mipre}.
In particular, the results from Ref.~\cite{ji2022mipre} ensure the existence of a nonlocal game (a particular case of a Bell-like inequality), such that correlations from $C_{qc}$ can win with unit probability, but correlations from $C_{qa}$ cannot attain a value greater than $1/2$.
\end{remark}

\begin{result}
By combining the results above, we see that there exist finite sets $A$, $B$, $X$, $Y$, such that for their associated scenario, we have the strict inclusion
\begin{equation}
C_q \subsetneq C_{qs} \subsetneq C_{qa} \subsetneq C_{qc}.
\end{equation} 
\label{mainresult}
\end{result}


\section{Possible consequences for physics}



\begin{figure}
\centering
\includegraphics[width=\linewidth]{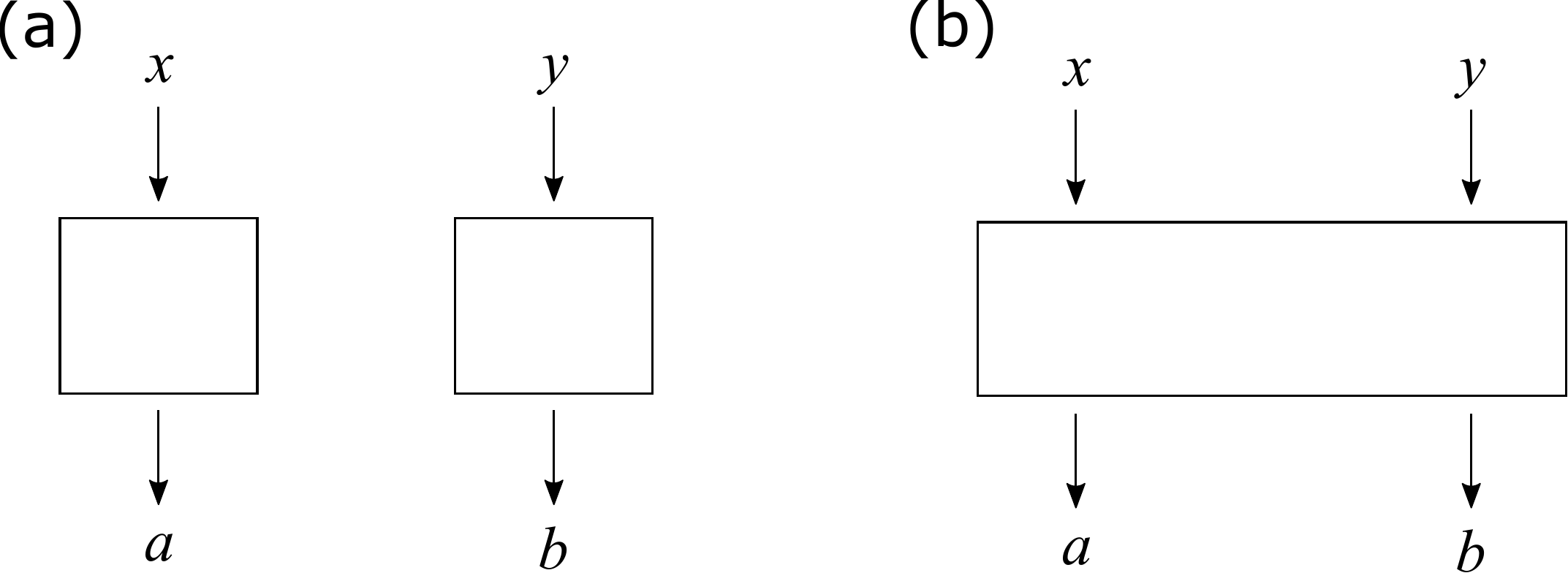}
\caption{(a)~Experiment in which each of two independent observers, Alice and Bob, performs a measurement on one of two spatially separated systems. (b)~Experiment in which Alice and Bob perform commuting measurements  on a single system. In both cases, $x$ ($y$) is Alice's (Bob's) measurement and $a$ ($b$) is the corresponding outcome.}
\label{fig1}
\end{figure}


The mathematical fact that $C_{qa}$ and $C_{qc}$ can be separated by a Bell-like inequality, hereafter called a \emph{quantum tensor inequality}, and that QM and AQFT allow, in principle, the existence of $p \in C_{qc}\setminus C_{qa}$ opens the path to experiments that detect violations of quantum tensor inequalities. However, it is not yet clear whether such experiments can be preformed, even in principle, and what their outcome would be.

\begin{figure*}
\centering
\includegraphics[width=\linewidth]{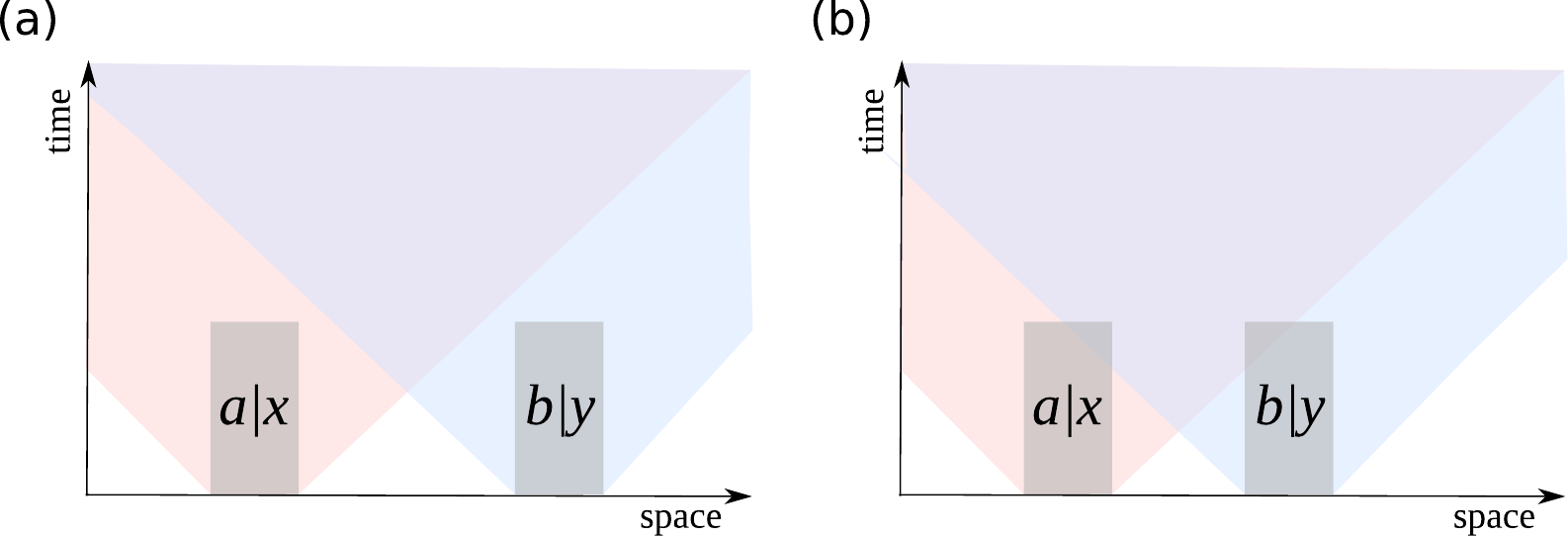}
\caption{Space-time diagram of an implementation of measurements $x$ and $y$ with respective outcomes $a$ and $b$. Each measurement requires a certain space-time region from the point where the measurement is decided till its outcome is recorded, an outer approximation of which is indicated by a gray box. For having spacelike separation, the implementation of measurement $y$ with outcome $b$ has to be outside of the forward light cone of any point of the implementation of measurement $x$ with outcome $a$, and vice versa. In (a), $a|x$ and $b|y$ are spacelike separated.
In (b), part of $a|x$ is in the light cone of $b|y$, and part of $b|y$ is in the light cone of $a|x$.}
\label{fig2}
\end{figure*}

Our objective here is to identify the logical possibilities depending on which experiments are feasible and what are their results.
For that, we distinguish between experiments in which there is a spatial separation between the systems, as in Fig.~\ref{fig1}~(a), and those commuting measurements performed on a single system, as in Fig.~\ref{fig1}~(b). In the first case, we also distinguish between experiments in which events are spacelike separated, as in Fig.~\ref{fig2}~(a), and experiments in which events are not spacelike separated, as in Fig.~\ref{fig2}~(b).

\begin{possibility}\label{p1}
No $p\in C_{qc}\setminus C_{qa}$ is feasible, even in experiments with commuting measurements on a single system.
\end{possibility}

\begin{possibility}\label{p2}
No $p\in C_{qc}\setminus C_{qa}$ is feasible in experiments with two spatially separated systems, but $p\in C_{qc}\setminus C_{qa}$ is feasible in experiments with commuting measurements on a single system.
\end{possibility}

\begin{possibility}\label{p3}
$p\in C_{qc}\setminus C_{qa}$ is feasible in experiments with two spatially separated systems, but only if the events are not spacelike separated.
\end{possibility}

\begin{possibility}\label{p4}
$p\in C_{qc}\setminus C_{qa}$ is feasible in experiments with two spatially separated systems, even if the events are spacelike separated.
\end{possibility}

Possibility~2 would mean that there are correlations on single systems that are impossible to achieve on spatially separated systems. In other words, that, for scenarios with identical joint measurability relations between the observables, there are Kochen-Specker contextual correlations (between non-disturbing measurements of jointly measurable observables) that are ``larger'' than any spatially separated correlation.

Possibility~3 would mean that there are correlations on composite systems which, if the events are not spacelike separated, would violate a quantum tensor inequality, while the violation would vanish whenever the events are spacelike separated.

Any of the possibilities 2, 3, and 4 would allow to produce correlations $p\in C_{qc}\setminus C_{qa}$ under specific conditions and hence open a path to experimentally prove the existence of systems with infinitely many degrees of freedom.

Correlations $p\in C_{qc}\setminus C_{qa}$ can fail to be feasible because of various obstacles, including:
\begin{itemize}
\item[(a)] A representation of the state and measurements leading to $p$ is not possible within accepted frameworks of QM/AQFT (see, for example, Ref.~\cite{Borsten2021PRD}).
\item[(b)] The existence of devices implementing the states and measurements leading to $p$ contradict some fundamental practical limitations (for example, having energy or time requirements not accessible due to cosmological constraints).
\item[(c)] The devices cannot be realized because of fundamental laws (for example, ideal measurements of sharp observables may require either infinite time or infinite energy, see, for example, Refs.~\cite{Hall2013,Guryanova:2020Q}.
\item[(d)] It is impossible to perform measurements on infinite-dimensional systems which do not correspond to measurements on finite-dimensional systems (note, for example, Refs.~\cite{Santos2005PRL,Santos:07}). 
\end{itemize}

In spite of all the possible obstacles, we believe it is important to design a strategy to try to know which of the four possibilities 1--4 occurs and to identify why.


\section{Roadmap for experimental tests}


\begin{problem}\label{prob1}
Which of the possibilities 1--4 is applicable in physics?
\end{problem}

A way to address Problem~\ref{prob1} would be to solve the following problems:

\begin{problem}\label{prob2}
Identify explicit scenarios in which there is $p\in C_{qc}\setminus C_{qa}$.
Here, by scenario we do not mean a Bell scenario, but a Kochen-Specker contextuality scenario with sharp observables whose relations of joint measurability are the same as those of a Bell scenario.
As far as we know, these scenarios could even include the ones that have the same relations of joint measurability than the $(3,3,2,2)$ \cite{Pal:2010PRA} or $(4,5,3,3)$ \cite{Coladangelo:2020NC} Bell scenarios.
\end{problem}

\begin{problem}\label{prob3}
For any of the scenarios obtained solving Problem~\ref{prob2}, identify explicit $p\in C_{qc}\setminus C_{qa}$.
\end{problem}

The results by Ji \emph{et al.} \cite{ji2022mipre, Ji:2021CACM} suggest that Problems~\ref{prob2} and~\ref{prob3} can be solved.

\begin{problem}\label{prob4}
Check whether the correlations obtained by solving Problem~\ref{prob3} have a representation in QM/AQFT. Consider obstacles to their physical implementation.
\end{problem}

\begin{problem}\label{prob5}
If there are no obstacles, identify quantum tensor inequalities violated by $p$ and find violations that can be tested in realistic (imperfect) experiments.
\end{problem}

\begin{remark}
The Navascués-Pironio-Acín hierarchy \cite{NPA_PRL, NPA_NJP} is a family of outer approximation that converges to $C_{qc}$. However, we do not have any hierarchy of inner approximations that converges to $C_{qc}$. Therefore, the $\varepsilon$-weak membership problem for $C_{qc}$ might be undecidable, as it is the case for $C_{qa}$ \cite{Ji:2021CACM}, see Result~5 and Refs.~\cite{ji2022mipre, Ji:2021CACM}.
However, even if it is undecidable, finding a quantum tensor inequality only requires finding an upper bound of a functional $T$ for any element in $C_{qa}$
which is below the maximum of $T$ for $C_{qc}$.
\end{remark}

So far, we have only considered the bipartite case. However, a more practical way to investigate the physics of correlations could be to extend $C_{qs}$, $C_{qc}$, and Problems~\ref{prob1}--\ref{prob5} to the multipartite case. From this perspective, it would be helpful to consider the following.

\begin{problem}\label{prob6}
How does the gap between tensor and commuting correlations grow as the number of systems grows?
\end{problem}

\section{Acknowledgments}


This work results from a session on ``Physical consequences of $\text{MIP}^\ast=\text{RE}$''
celebrated in the Centro de Ciencias Pedro Pascual in Benasque (Spain) on June 22, 2023 within the workshop ``Quantum Information''. This updated version of the manuscript was completed in the 2025 edition of  workshop ``Quantum Information''.
We thank all the participants for their feedback, in special Zoltán Zimborás for useful insights and fruitful discussions.
We also thank Andrea Coladangelo and Marcelo Fran\c{c}a Santos for additional help.
This work was supported by
AEI/MICINN (Project No.~PID2020-113738GB-I00),
the Canada-EU project ``Foundations of Quantum Computational Advantage'' (FoQaCiA) (doi: 10.3030/101070558),
the Deutsche Forschungsgemeinschaft (DFG, German Research Foundation, Project No.~447948357 and No.~440958198),
the Sino-German Center for Research Promotion (Project No.~M-0294),
the ERC (Consolidator Grant~683107/TempoQ), and
the German Ministry of Education and Research (Project QuKuK, BMBF Grant No.~16KIS1618K).


\bibliographystyle{quantum}

\bibliography{used}


\end{document}